\documentclass[
 reprint,
superscriptaddress,
showpacs,preprintnumbers,
 amsmath,amssymb,
 aps,
 prl, twocolumn
]{revtex4}

\usepackage{graphicx}% Include figure files
\usepackage{dcolumn}% Align table columns on decimal point
\usepackage{bm}% bold math
\usepackage{hyperref}% add hypertext capabilities
\usepackage{color}
\begin{document}

%\preprint{APS/123-QED}

\title{Bulk Fermi surface and electronic properties of Cu$_{0.07}$Bi$_{2}$Se$_{3}$}

\author{C. Martin}
\affiliation{Department of Physics, University of Florida, Gainesville, Florida 32611, USA}
\author{V. Craciun}
\affiliation{Laser Department, National Institute for Lasers, Plasma, and Radiation Physics, Magurele, Romania}
\author{K. H. Miller}
\affiliation{Department of Physics, University of Florida, Gainesville, Florida 32611, USA}
\author{B. Uzakbaiuly}
\affiliation{Department of Physics, University of Florida, Gainesville, Florida 32611, USA}
\author{S. Buvaev}
\affiliation{Department of Physics, University of Florida, Gainesville, Florida 32611, USA}
\author{H. Berger}
\affiliation{Institute of Physics of Complex Matter, Ecole Polytechnique Federal de Lausanne, CH-1015 Lausanne, Switzerland}
\author{A. F. Hebard}
\affiliation{Department of Physics, University of Florida, Gainesville, Florida 32611, USA}
\author{D. B. Tanner}
\affiliation{Department of Physics, University of Florida, Gainesville, Florida 32611, USA}

\date{\today}

\begin{abstract}
The electronic properties of Cu$_{0.07}$Bi$_{2}$Se$_{3}$ have been investigated using Shubnikov-de Haas and optical reflectance measurements. Quantum oscillations reveal a bulk, three-dimensional Fermi surface with anisotropy $k^{c}_{F}/k^{ab}_{F}\approx$ 2 and a modest increase in free-carrier concentration and in scattering rate with respect to the undoped Bi$_{2}$Se$_{3}$, also confirmed by reflectivity data. The effective mass is almost identical to that of Bi$_{2}$Se$_{3}$. Optical conductivity reveals a strong enhancement of the bound impurity bands with Cu addition, suggesting that a significant number of Cu atoms enter the interstitial sites between Bi and Se layers or may even substitute for Bi. This conclusion is also supported by X-ray diffraction measurements, where a significant increase of microstrain was found in Cu$_{0.07}$Bi$_{2}$Se$_{3}$, compared to Bi$_{2}$Se$_{3}$. 

\end{abstract}
\maketitle
Topological insulators (TI) have attracted considerable attention for the past several years, as the surface states, with Dirac band dispersion and protected by time-reversal symmetry, may be explored for spin-related applications\cite{Hasan10}. Bi$_{2}$Se$_{3}$ represents a particular  example of the topological insulator, where the Dirac states emerge at the surface of a three-dimensional bulk band structure\cite{Xia09}. Moreover, the Cu-doped material, Cu$_{x}$Bi$_{2}$Se$_{3}$, is superconducting for $x\geq 0.1$, with maximum $T_{c}\approx 4$ K\cite{Hor10}, opening the possibility for exploring other novel phenomena, like magnetic monopole and Majorana fermions\cite{Qi11, Sasaki11}.

It is important to understand the effect of Cu addition on the electronic structure, and further, if possible, on the origin of superconductivity in Cu$_{x}$Bi$_{2}$Se$_{3}$. A combined tunneling and photoemission study\cite{Wang11} found at least three different ways in which Cu atoms enter into the crystal structure of Bi$_{2}$Se$_{3}$: they can be adsorbed on the Se surface, occupy interstitial sites between the Bi and Se planes, or they can intercalate into the van der Waals gaps between the quintuple layered Se-Bi-Se-Bi-Se. While the adsorbed Cu is expected to affect only the surface properties, both the interstitial and intercalated atoms were found to act as donors, thus adding conduction electrons into the bulk. Moreover, there is also the possibility that Cu substitutes for Bi, in which case it acts as an acceptor, i.e. a hole dopant. Therefore, the bulk carrier concentration $n$ is expected to vary nonlinearly with $x$ in Cu$_{x}$Bi$_{2}$Se$_{3}$. This behavior has been confirmed by angle resolved photoemission spectroscopy (ARPES)~\cite{Wray10, Wray11}. 

In nominally undoped crystals of Bi$_{2}$Se$_{3}$, $n$ varies over several orders of magnitude, from $\sim 10^{16}$~cm$^{-3}$ to $\sim 10^{19}$~cm$^{-3}$. With Cu addition, superconductivity seems to appear when $n\geq5\times10^{19}$~cm$^{-3}$. Heat capacity measurements\cite{Kriener11} on superconducting Cu$_{x}$Bi$_{2}$Se$_{3}$, with $x=0.29$ and $n=1.3\times10^{20}$~cm$^{-3}$, found a manyfold enhancement of the electron effective mass, with $m^{*} = 2.6m_{0}$, compared to undoped Bi$_{2}$Se$_{3}$, where $m^{*}\approx~0.14m_{0}$~\cite{Butch10, Analytis10}. However, a much less significant increase, $m^{*}\approx~0.2m_{0}$, was observed from quantum oscillations for similar doping level, $x=0.25$ and $n=4.3\times10^{19}$~cm$^{-3}$~\cite{Lawson12}. In both Ref.~\cite{Kriener11} and Ref.~\cite{Lawson12}. the critical temperature was very similar $T_{c}\approx3$~K, underlying the complex nature of doping in Bi$_{2}$Se$_{3}$.

Here we present a study of X-ray diffraction, elemental analysis, quantum oscillations and optical spectroscopy of Cu$_{x}$Bi$_{2}$Se$_{3}$ with $x=0.07$, just below the threshold for superconductivity. We focus on the effect of Cu addition on the band structure and transport properties; these may be relevant to the emergence of superconductivity.

Single crystals of Cu$_{0.07}$Bi$_{2}$Se$_{3}$ were grown from high purity Bi, Se and Cu elements by the Bridgman technique. For comparison, single crystals of undoped Bi$_{2}$Se$_{3}$ were also synthesized under the same conditions and used in the present study. Optical reflectance measurements, using a Bruker 113v spectrometer, were performed on two samples of Cu$_{0.07}$Bi$_{2}$Se$_{3}$ and one of Bi$_{2}$Se$_{3}$. The crystals had approximate dimensions 6$\times$4$\times$0.150 mm$^{3}$,. The results on the Cu doped samples were nearly identical and one was further used for Shubnikov-de Haas oscillations and the other, together with the Bi$_{2}$Se$_{3}$ sample, were investigated by X-ray diffraction (XRD) and energy dispersive X-ray analysis (EDX). Longitudinal and Hall resistance were measured in the SCM-2 facility at the National High Magnetic Field in Tallahassee. The facility consists of a top loading $^{3}$He cryostat, with sample in liquid and a base temperature of 0.3 K in conjunction with an 18-20 Tesla superconducting magnet. A rotation probe with an angular resolution better than 1$^{\circ}$ was used for angle dependent measurements. Electrical contact was made by attaching gold wires with silver paint. XRD data was taken using a Panalytical X'Pert MRD diffractometer and for EDX a Philips XL40 FEG-SEM was used.

\begin{figure}[htp]
\includegraphics[width=0.4\textwidth]{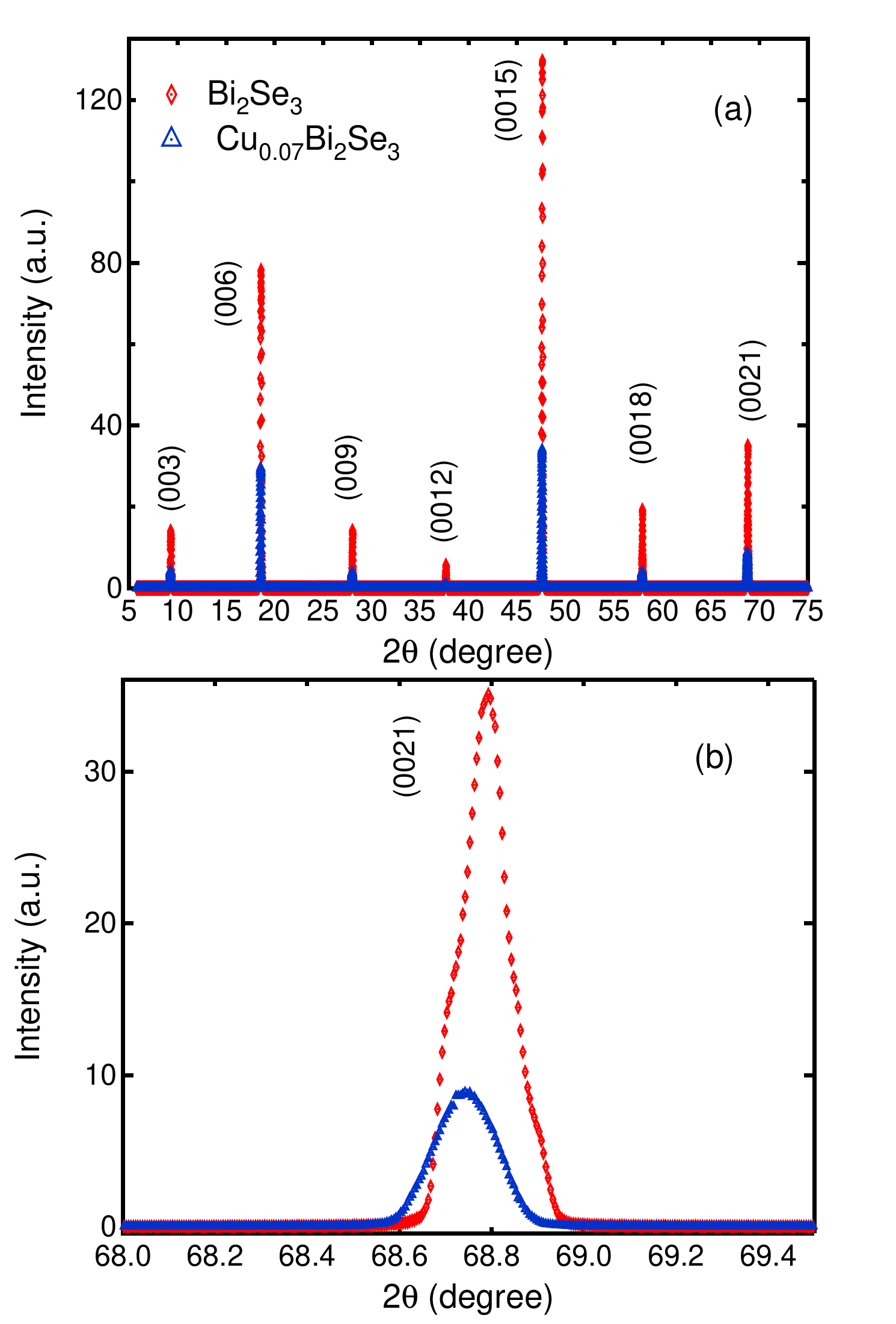}                
 \caption{(Color online) (a) $X$-ray diffraction pattern of Bi$_{2}$Se$_{3}$ and Cu$_{0.07}$Bi$_{2}$Se$_{3}$, respectively. (b) The $(0021)$ peak}.  
\label{Fig1}
\end{figure}

The stoichiometry of samples and the presence of Cu were confirmed by EDX analysis. For a 7\% nominal Cu composition, EDX data on Cu$_{x}$Bi$_{2}$Se$_{3}$ yields $x = 0.061~\pm~0.010$, consistent within error with the nominal value. Figure~\ref{Fig1}(a) shows the X-ray diffraction of single crystals of Cu$_{0.07}$Bi$_{2}$Se$_{3}$ and Bi$_{2}$Se$_{3}$. Only sharp reflections from the $(00l)$ planes, with no additional signal from other phases are visible. A slight increase with Cu doping was observed in the $c$-axis, from $c =28.649$~\AA\ to $c=28.662$~\AA. Line profile analysis reveals significant broadening and distortion of the diffraction peaks with Cu doping, indicating the presence of defects and/or microstrains. An example of the $(0021)$ peak is shown in Fig.~\ref{Fig1}(b). The FWHM of Bi$_{2}$Se$_{3}$ is $0.120^{\circ}$, whereas in the Cu$_{0.07}$Bi$_{2}$Se$_{3}$ it is $0.166^{\circ}$, almost 40\% larger. From Williamson-Hall plots we found that in Cu$_{0.07}$Bi$_{2}$Se$_{3}$ the crystallite size decreases from 94 nm to 79 nm and the microstrain increases significantly, from 0.003\% to 0.010\%. The very small change in $c$-axis lattice constant and the defects and strain found in Cu$_{0.07}$Bi$_{2}$Se$_{3}$ suggest that a significant amount of Cu must enter interstitially or substitute for Bi, rather then in the van der Waals gaps. This conclusion will be supported by our further experimental findings.

\begin{figure}[htp]
\includegraphics[width=0.4\textwidth]{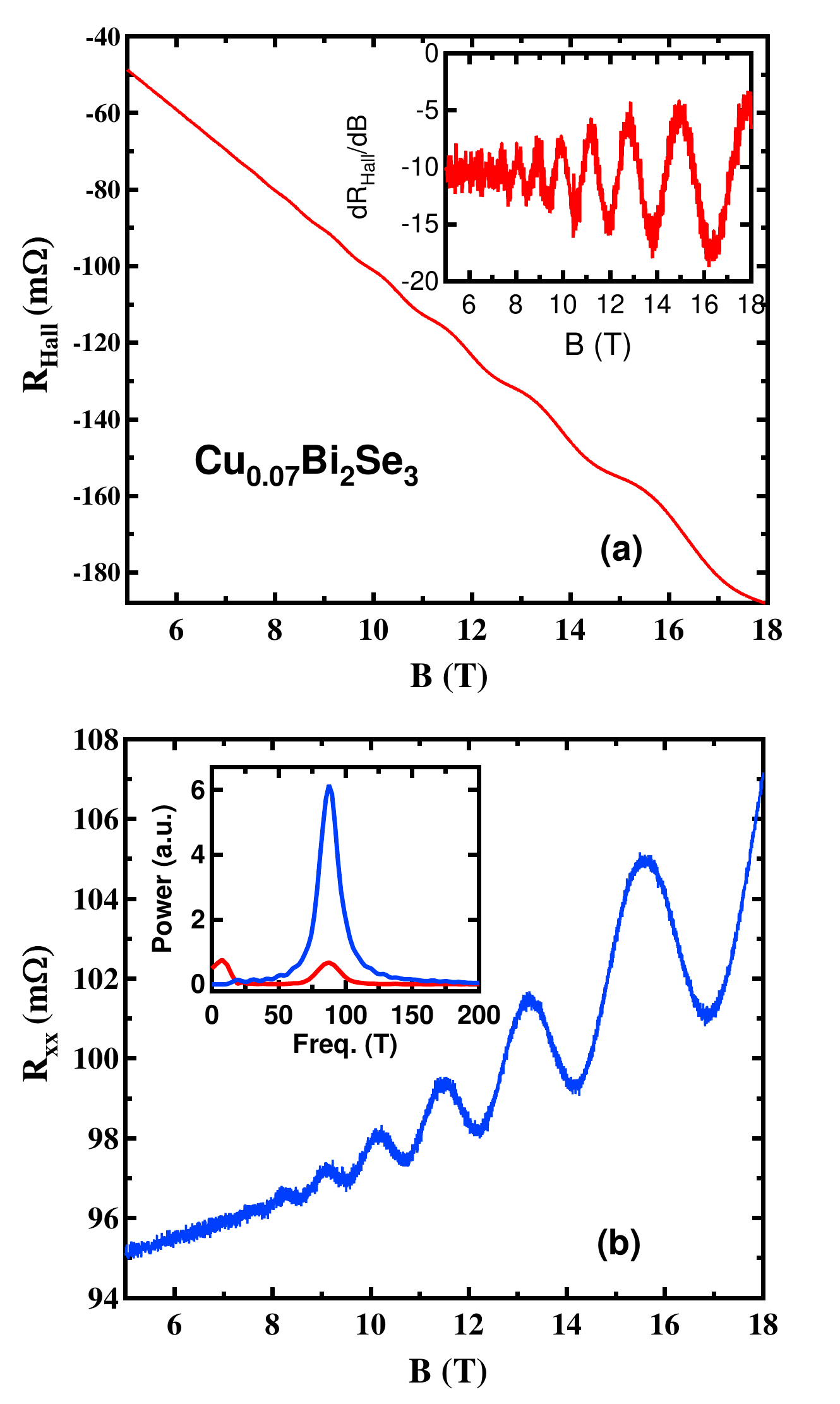}                
 \caption{(Color online) (a) Magnetic field dependence of the Hall resistance at $T = 0.3$~K and above 5~T, to emphasize the oscillatory behavior. (b) In-plane resistance $R_{xx}$ for the same temperature and the same magnetic field range as (a). Inset (a) shows $dR_{H}/dH$. Inset (b) shows the Fourier transform of  $R_{xx}(1/H)$ (blue) and $R_{H}(1/H)$ (red), respectively.} 
\label{Fig2}
\end{figure}

\begin{figure}[htp]
\includegraphics[width=0.4\textwidth]{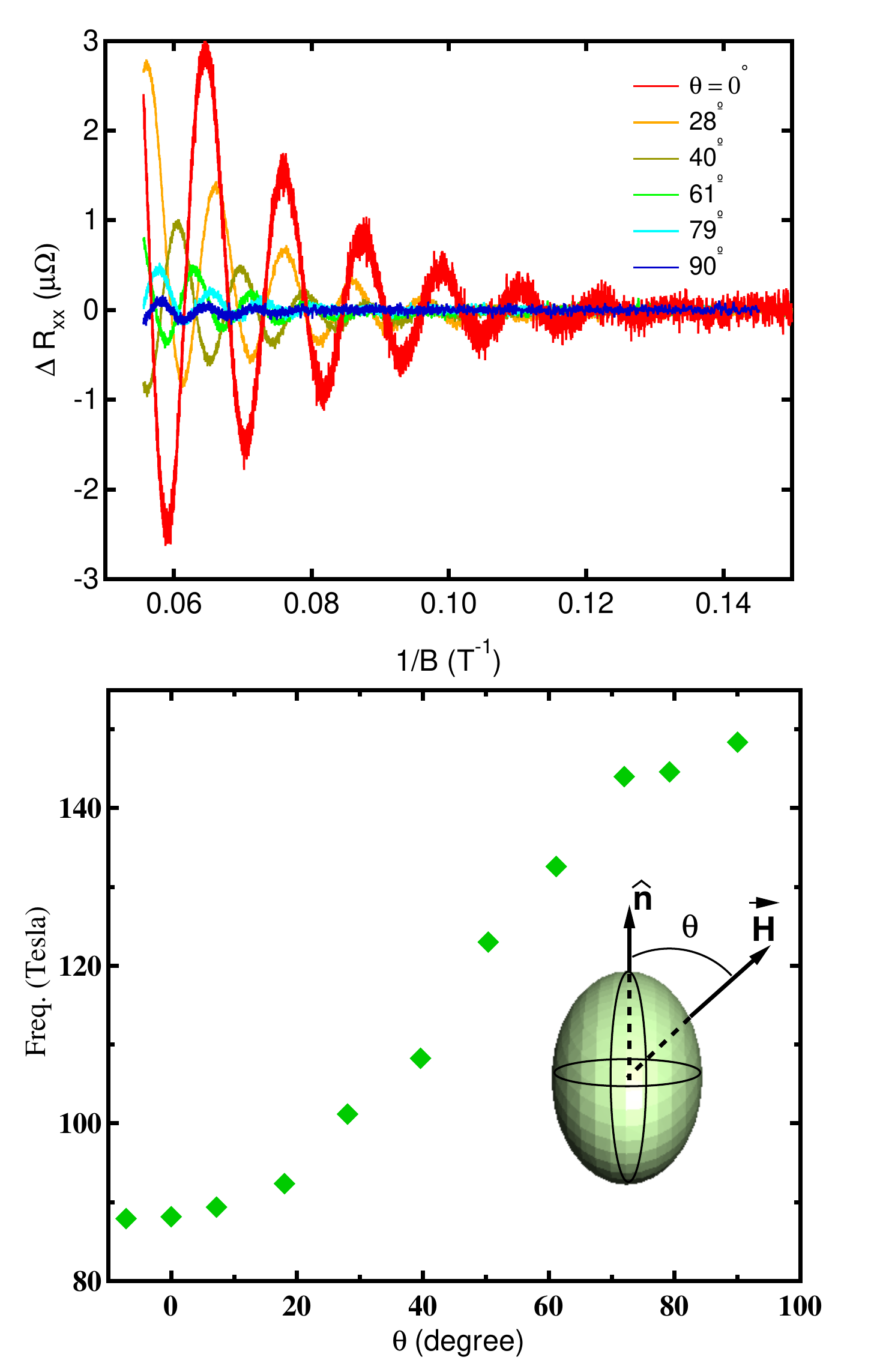}                
 \caption{(Color online) (a) $R_{xx}$~vs.~$1/H$ at $T = 0.3$ K for different orientations of the magnetic field with respect to the $c$-axis. (b) Angle dependence of the oscillation frequency. The inset shows a sketch of a 3D ellipsoid Fermi pocket and the orientation of the applied magnetic field.}  
\label{Fig3}
\end{figure}

Figure~\ref{Fig2} shows the high magnetic field Hall resistance (Fig.~\ref{Fig2}(a)) and longitudinal resistance (Fig.~\ref{Fig2}(b)) at $T=0.3$~K, with the magnetic field applied along the crystallographic $c$-axis. Clear oscillations can be observed in both traces. Although their amplitude is much smaller in the Hall signal, they appear very clearly in the inset of Fig.~\ref{Fig2}(a), where we plot $dR_{H}/dH$. Fourier transform (FFT) of $R_{xx}(1/H)$ and $R_{H}(1/H)$ yields $F = 87.4 \pm 0.1$~Tesla (inset of Fig.~\ref{Fig2}(b)). We rotated the sample in the magnetic field and observed  oscillations for all orientations. Figure~\ref{Fig3}(a) shows the oscillations at several angles, indicating the 3-D nature of the Fermi surface. The angular dependence of the frequency is plotted in Fig.~\ref{Fig3}(b). The frequency when $B\perp c$ is $F = 146 \pm 2$~Tesla. Assuming an ellipsoidal Fermi pocket, the corresponding momenta are  $k^{ab}_{F} = 0.5$~nm$^{-1}$ in the $ab$-plane and $k^{c}_{F} = 0.9$~nm$^{-1}$ along the $c$-axis, respectively. The anisotropy is $k^{c}_{F}/k^{ab}_{F} = 1.8$, slightly larger than previous values of $k^{c}_{F}/k^{ab}_{F}=1.2$--1.6 obtained for Bi$_{2}$Se$_{3}$~\cite{Butch10, Analytis10}, making unlikely that anisotropy plays an important role in superconductivity. Moreover, anisotropy was found to decrease slightly for 25\% Cu doping\cite{Lawson12}. These results are rather relevant for the non-monotonic effect of Cu addition on the band structure of Bi$_{2}$Se$_{3}$. The carrier concentration is $n=\left(1/3\pi^{2}\right)k^{ab^{2}}_{F}k^{c}_{F}\approx 5\times10^{18}~$cm$^{-3}$, almost identical to the value we obtained from Hall coefficient. It is important to mention that larger carrier concentrations, up to about 2--3 $\times$10$^{19} $~cm$^{-3}$ were reported in recent studies of undoped Bi$_{2}$Se$_{3}$\cite{Analytis10, Lawson12} and that the superconducting samples of Cu$_{x}$Bi$_{2}$Se$_{3}$ seem to have $n\geq 5\times 10^{19}$~cm$^{-3}$. Therefore, it is possible that there is a threshold carrier concentration required for superconductivity.

\begin{figure}[htp]
\includegraphics[width=0.5\textwidth]{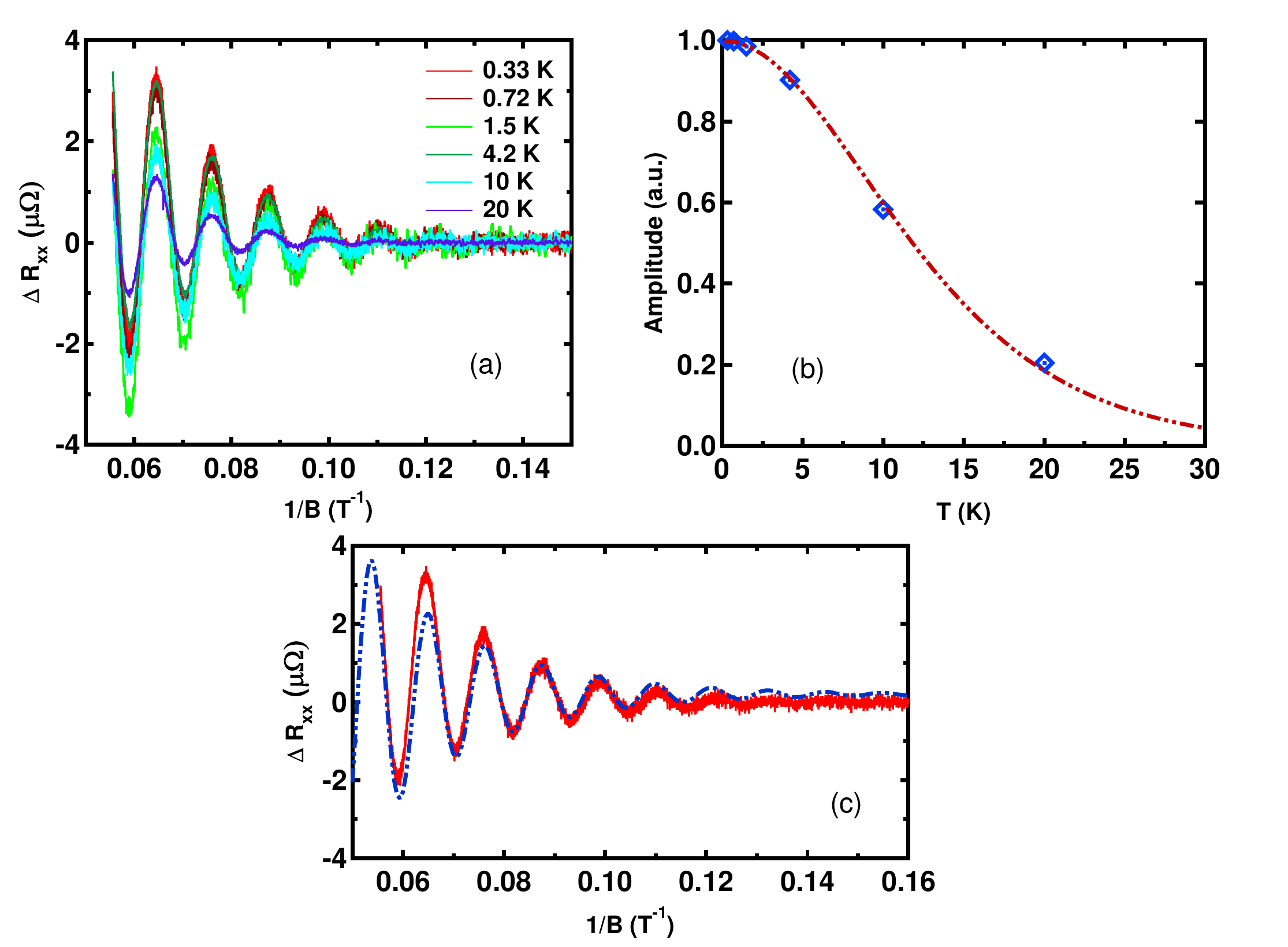}                
\caption{(Color online) (a) $R_{xx}$~vs.~$1/H$ at different temperatures for the magnetic field along the ${c}$-axis. (b) Temperature dependence of the FFT amplitude (symbols), normalized to the base (0.3 K) temperature, and a fit to the Lifshitz-Kosevich formula, as explained in the main text. (c) $R_{xx}(1/H)$ at $T = 0.3$~K (continuous red line) and a fit to the field dependence expression explained in the main text (continuous-dashed blue line).}
\label{Fig4}
\end{figure}

Another important parameter is the carrier effective mass, given in particular the large value reported in Ref.~\cite{Kriener11} for $x = 0.29$. Figure~\ref{Fig4}(a) displays the oscillatory part (after second order polynomial background subtraction) of $R _{xx}$ versus inverse magnetic field at different temperatures. Clear oscillations can be observed up to at least T=20K and in Fig.~\ref{Fig4}(b), the amplitude of the FFT with temperature is fitted to the Lifshitz-Kosevich function: $\gamma T/\sinh(\gamma T)$, with $\gamma = 14.69 m^{*}/m_{0} B$, where $B$ is the magnetic field, $m^{*}$ is the effective mass, and $m_{0}$ the rest mass of the electron\cite{Shoenberg84}. The same analysis was also performed on the Hall resistance data and we obtain an average effective mass  $m^{*} =0.140\pm0.006$. Surprisingly, this value is nearly identical to those previously obtained on the undoped Bi$_{2}$Se$_{3}$\cite{Butch10, Analytis10}. Nevertheless, our result for 7\% Cu doping seems to be in line with a recent report, where a very small enhancement ($m^{*} =0.19m_{0}$) was reported for the superconducting samples with $x = 0.25$, three times as much Cu as in our samples\cite{Lawson12}. Therefore, while a large increase of $m^{*}$ may occur with Cu doping in some cases, it does not appear to be a general trend.

From the field dependence of SdH oscillations we extract the scattering rate. The amplitude of oscillations increases with magnetic field as $\Delta R_{xx}\propto\exp(-\gamma T_{D})\cos(2\pi F/B+\pi)$, where $\gamma$ was defined earlier and $T_{D} = \hbar/\left(2\pi\tau k_{B}\right)$ is the Dingle temperature, related to the lifetime $\tau$ of the electrons\cite{Shoenberg84}. From the fit to the expression above for $\overrightarrow{H}\|\hat{c}$-axis, shown in Fig.~\ref{Fig4}(c), we extract $T_{D} = 26\pm1.5$ K and hence, the scattering time $\tau = (4.7 \pm 0.3)\times 10^{-14}$ s (or rate $1/\tau = 114 \pm 73$ cm$^{-1}$). Our result for the Dingle temperature is larger than previous reports of $T_{D}$ = 4--24 K\cite{Butch10, Analytis10, Lawson12}, reflecting probably the microstrain and the distortion observed in the X-ray data. We note that in a recent study,\cite{Kriener12}
superconducting samples with large volume fraction and higher
transition temperatues had mean free paths in the range of $\ell =
10$--40 nm. We calculate for our sample that $\ell = \hbar k_{F}\tau/m^{*}\approx 20$~nm. Therefore, the mean free path in our non-superconducting samples is of the same order as those that are superconducting.

\begin{figure}[htp]
\includegraphics[width=0.4\textwidth]{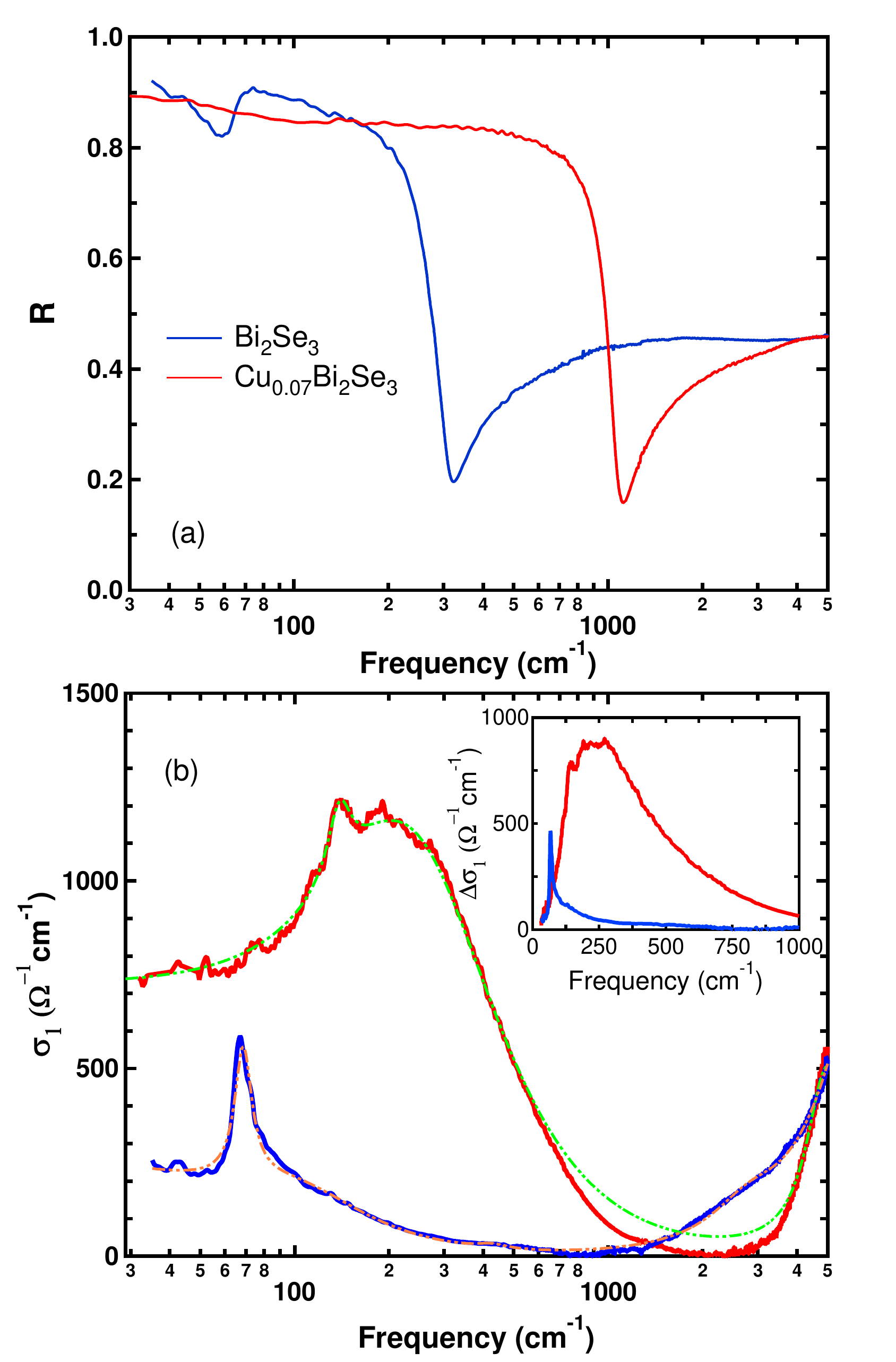}                
 \caption{(Color online) (a) Room temperature reflectance of Bi$_{2}$Se$_{3}$ and Cu$_{0.07}$Bi$_{2}$Se$_{3}$, respectively. (b) Optical conductivity of Bi$_{2}$Se$_{3}$ and Cu$_{0.07}$Bi$_{2}$Se$_{3}$, respectively. The dash-dotted green lines in (b) are Lorentz-Drude fits to the data. The inset of (b) represents the Drude (free carrier) conductivity calculated from the fitting parameters.}  
\label{Fig5}
\end{figure}

We also studied the effect of Cu substitution in Bi$_{2}$Se$_{3}$ using optical reflectance measurements. Fig.~\ref{Fig5} compares the room temperature optical reflectance $\cal R(\omega)$ for undoped Bi$_{2}$Se$_{3}$ and Cu$_{0.07}$Bi$_{2}$Se$_{3}$. It can be clearly observed that the free-carrier plasma edge occurs at higher frequency in the Cu$_{0.07}$Bi$_{2}$Se$_{3}$ sample. Considering that we found that the effective mass in both samples is very similar, this is consistent with an increase in free-carrier concentration as discussed below. We also notice that the strong phonon mode near 60 cm$^{-1}$ is almost completely suppressed with Cu addition, whereas the one near 130 cm$^{-1}$, barely observable in Bi$_{2}$Se$_{3}$ at room temperature, appears enhanced in Cu$_{0.07}$Bi$_{2}$Se$_{3}$. The vibrational modes in Bi-based topological insulators have been studied in detail elsewhere\cite{Richter77, LaForge10, Pietro12}; here, we restrict our discussion to the effect of Cu on the electronic structure.

Figure~\ref{Fig5}(a) shows the real part of the optical conductivity $\sigma_{1}(\omega)$ obtained after Kramers-Kronig transformation of $\cal R(\omega)$~\cite{Wooten72}. One can easily notice the remarkable increase in the far-infrared spectral weight with Cu addition, but we will show shortly that this is not solely from a free-carrier (Drude) contribution. From fit to a Lorentz-Drude model, we found that the Drude plasma frequency increases from $\omega_{p}^{D}=\sqrt{ne^{2}/m^{*}\epsilon_{0}}\approx 1000$ cm$^{-1}$ in Bi$_{2}$Se$_{3}$ to  2100 cm$^{-1}$ in Cu$_{0.07}$Bi$_{2}$Se$_{3}$. The scattering rate was also found to increase, from $1/\tau\approx 60$ cm$^{-1}$ to 100 cm$^{-1}$, respectively. Considering the effective mass $m^{*}=0.14 m_{0}$ determined from SdH oscillations, the free carrier concentration obtained from optical conductivity is $1.6\times 10^{18}$ cm$^{-3}$ in Bi$_{2}$Se$_{3}$ and 7$\times$10$^{18}$cm$^{-3}$ in Cu$_{0.07}$Bi$_{2}$Se$_{3}$. The latter value is in fair agreement with that obtained above from SdH oscillations and would imply 
that Cu addition gives a fourfold enhancement in carrier concentration. However, this is much less than the possible variations that may result from stoichiometric variations occurring in the crystal growth process. 
\begin{figure}[htp]
\includegraphics[width=0.4\textwidth]{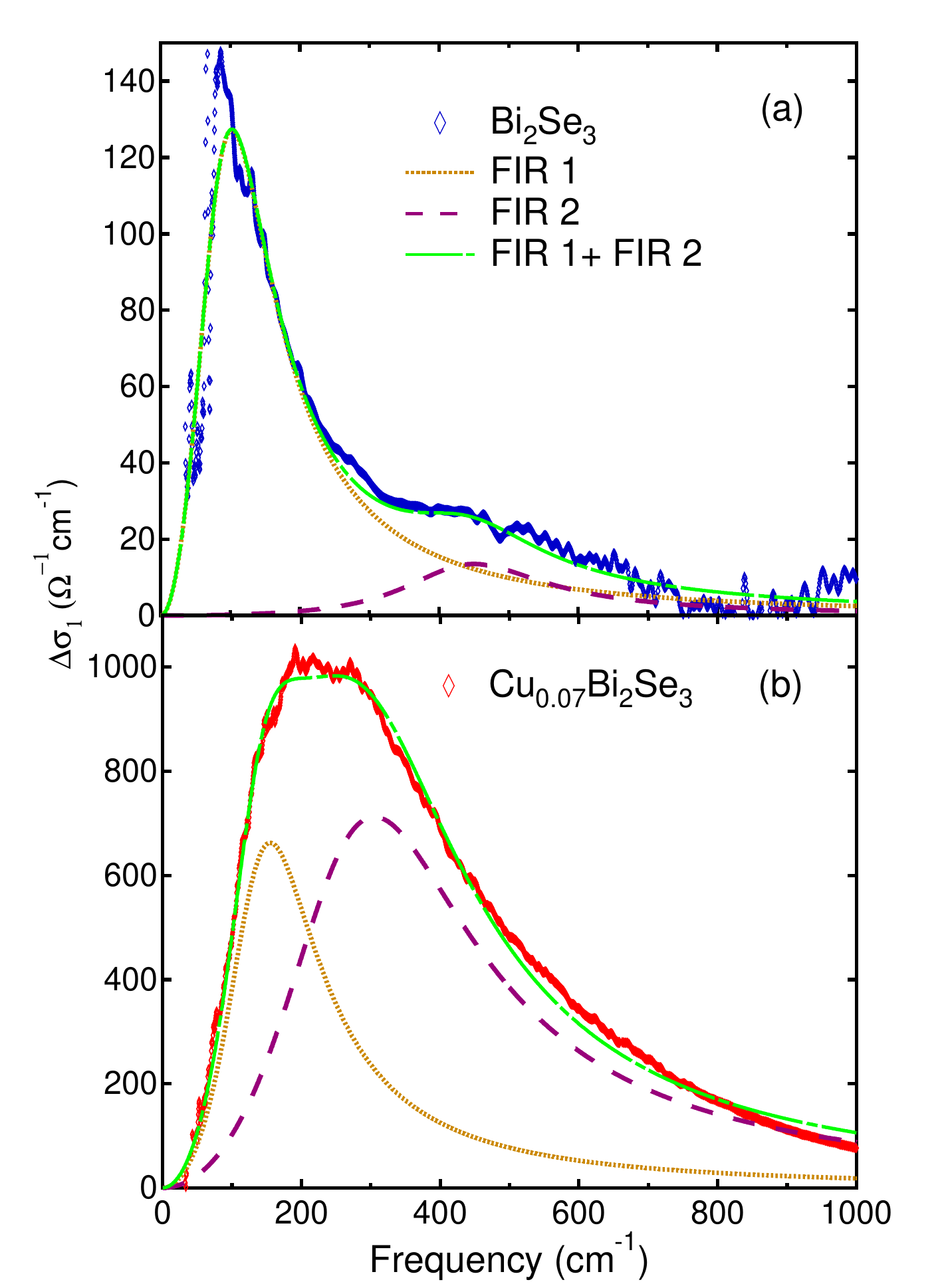}                
 \caption{(Color online) Room temperature optical conductivity of Bi$_{2}$Se$_{3}$ (a) and Cu$_{0.07}$Bi$_{2}$Se$_{3}$ (b), obtained after subtracting the Drude and the sharp phonons contributions.}  
\label{Fig6}
\end{figure}

To emphasize that there is another strong effect of Cu doping in addition to the increased free carrier concentration, we plot in Fig.~\ref{Fig5} the remaining optical conductivity $\Delta\sigma_{1}(\omega)$ obtained after subtracting the Drude and the sharp phonon contributions. The data are similar to a very recent study on several Bi-based TIs\cite{Pietro12}, and we describe it the same way, in terms of two absorption bands: a low frequency FIR1, centered around 150 cm$^{-1}$, and a higher frequency one, FIR2 ($\approx$ 400 cm$^{-1}$). As was pointed out in Ref.~\cite{Pietro12}, they resemble the phosphorous (P) doped silicon (Si), where P impurities give rise to bound states, with the outer electrons in a hydrogen-like potential\cite{Thomas81}. The low frequency band, FIR1 in our case, would then correspond to hydrogen-like $1s\rightarrow np$ transitions and the higher frequency (FIR2) to impurity ionization, i.e. transition from the bound states to the continuum (conduction) band. What is striking in our Cu doped sample (Fig.~\ref{Fig5}(b)) is that the amplitude of the FIR2 peak is several times larger than that observed in any of the previous Bi-based TIs\cite{Pietro12}, reflecting a large concentration of bound electrons. From energy considerations, it is expected that the bound states are created by Cu atoms that enter interstitially, within the Bi-Se ionic bond, or even by Cu atoms substituting for Bi in the crystal structure. Localization of electrons from the Cu atoms intercalated into the van der Waals gaps between the quintuple layers would imply strong Coulomb repulsion and hence, possible magnetic order of Cu ions and enhancement of the effective mass due to electron-electron correlation. We have seen however from SdH oscillations that there is no change in $m^{*}$ and no magnetic correlations were ever reported in Cu$_{x}$Bi$_{2}$Se$_{3}$. Therefore, we interpret the strong impurity bands seen in Fig.~\ref{Fig5}(b) as arising from a large amount of Cu occupying interstitial sites between the Bi and Se layers, or even substituting for Bi atoms, just as we concluded from the X-ray analysis.

In conclusion, we performed combined measurements of X-ray diffraction, quantum oscillations, and optical spectroscopy on Cu$_{0.07}$Bi$_{2}$Se$_{3}$. Quantum oscillations reveal a bulk, 3-D Fermi surface, with the following properties: carrier concentration $n = (6 \pm 1) \times 10^{18}$~cm$^{-3}$, about four times larger than that of the undoped Bi$_{2}$Se$_{3}$; effective mass $m^{*}$ unchanged with doping and a larger scattering rate than in Bi$_{2}$Se$_{3}$. Given that only a slight increase of $m^{*}$ was observed in the superconducting samples with 25\% Cu\cite{Lawson12}, we suggest that the most important parameter for triggering superconductivity is the carrier concentration. In our samples this is significantly lower than the value of  $n\geq 5 \times 10^{19}$ cm$^{-3}$ measured for superconducting samples. Optical conductivity shows that, besides free carriers, Cu also gives rise to a remarkably strong hydrogen-like impurity bound state. Therefore, a significant amount of Cu must enter the interstitial space between the Bi and Se layers or even substitute for Bi, a fact also confirmed by the X-ray data. For further understanding, it would be useful to do an comparative optical study of the bound states between superconducting and non-superconducting samples of Cu$_{x}$Bi$_{2}$Se$_{3}$.      

Portions of this work were supported by the U.S. Department of Energy through contract No.~ DE-FG02-02ER45984 at the University of Florida and by the National Science Foundation grant DMR 1005301. The work of V. Craciun was funded by CNCSIS Ideas Project code 337. A portion was performed at the National High Magnetic Field Laboratory, which is supported by National Science Foundation Cooperative Agreement No. DMR-0654118, the State of Florida, and the U.S. Department of Energy. We would like to thank Ju-Hyun Park, Glover Jones, and Timothy Murphy for support with the experiment at the National High Magnetic Field Laboratory.

%\bibliographystyle{apsrev4-1}
%\bibliography{reference}% Produces the bibliography via BibTeX.  

\end{document}